\documentstyle[prb,aps,preprint,epsf]{revtex}
\newcommand{\sech}{{\rm sech}}
\newcommand{\hlf}{{\textstyle\frac{1}{2}}}
\newcommand{\qrtr}{{\textstyle\frac{1}{4}}}
\newcommand{\hc}{{\rm h.c.}}
\newcommand{\dg}{^{\dagger}}
\newcommand{\gdg}{^{\ }}     % `Ghost DaGger' for annihilation operators
\newcommand{\ket}[1]{|#1\rangle}
\newcommand{\bra}[1]{\langle #1|}
\newcommand{\up}{\uparrow}
\newcommand{\dn}{\downarrow}
\newcommand{\updn}{\ket{\!\up\!\dn}}
\newcommand{\bvec}[1]{\vec{#1}}
\newcommand{\svec}[1]{\vec{#1}}
\newcommand{\epsk}{\epsilon_{\svec{k}}}
\newcommand{\tepsk}{\tilde{\epsilon}_{\svec{k}}}
\newcommand{\nn}[1]{\langle #1 \rangle}
\newcommand{\nnij}{\nn{ij}}
\newcommand{\urs}{URu$_2$Si$_2$}
\newcommand{\eq}[1]{Eq.~(\ref{#1})}
\newcommand{\figref}[1]{Fig.~\ref{#1}}
\newcommand{\refcite}[1]{Ref.~[$\!\!$\onlinecite{#1}]}

\newcommand{\Elevel}{\begin{picture}(100,10)(0,0)\thinlines 
\put(0,4){\line(1,0){100}}
\put(0,6){\line(1,0){100}}
\end{picture}}
\newcommand{\StevensOp}[2]{\hat{O}_{#1}^{#2}}
\newcommand{\StevensCoeff}[2]{B_{#1}^{#2}}
\newcommand{\HCFterm}[2]{\StevensCoeff{#1}{#2}\StevensOp{#1}{#2}}
\newcommand{\Gam}[2]{|\Gamma^{#1}_{#2}\rangle}
\begin{document}
\title{The Ising-Kondo lattice with transverse field:\\
an \boldmath{$f$}-moment Hamiltonian for URu$_{\bf 2}$Si$_{\bf 2}$?}
\author{A.E.~Sikkema$^1$,
        W.J.L.~Buyers$^{2,3}$,
        I.~Affleck$^{1,2}$, \&
        J.~Gan$^4$}
\address{$^1$Department of Physics and $^2$Canadian Institute for Advanced
Research,\\
University of British Columbia, Vancouver, BC, Canada~~V6T~1Z1\\
$^3$AECL, Chalk River, ON, Canada~~K0J~1J0\\
$^4$Department of Physics, University of California, Berkeley, CA~94720\\
}
\date{Submitted to {\sl Phys. Rev. B}, 19 April 1996}

\maketitle
\abstract{
We study the phase diagram of the Ising-Kondo lattice with transverse
magnetic field as a possible model for the weak-moment heavy-fermion
compound \urs, in terms of two low-lying $f$ singlets in which the
uranium moment is coupled by on-site exchange to the conduction
electron spins.  In the mean-field approximation for an extended range
of parameters, we show that the conduction electron magnetization
responds logarithmically to $f$-moment formation, that the ordered
moment in the antiferromagnetic state is anomalously small, and that
the N\'eel temperature is of the order observed.  The model gives a
qualitatively correct temperature-dependence, but not magnitude, of
the specific heat.  The majority of the specific heat jump at the
N\'eel temperature arises from the formation of a spin gap in the
conduction electron spectrum.  We also discuss the single-impurity
version of the model and speculate on ways to increase the specific
heat coefficient.  In the limits of small bandwidth and of small
Ising-Kondo coupling, we find that the model corresponds to
anisotropic Heisenberg and Hubbard models respectively.
}

\section{Introduction}

As the large specific heat jump of \urs\ at its N\'eel temperature of
$T_N=17.5$~K is difficult to reconcile with its anomalously small
staggered magnetization\cite{Broholm87,Broholm91}, some
authors\cite{Ramirez,Barzykin,Cox} have speculated that there is some
hidden order parameter.  Recent neutron experiments indicate, however,
that the weak Bragg peaks of the ordered phase break time reversal
symmetry as would be the case if the order were magnetic
dipolar\cite{Walker93}.  In addition the hidden multispin order
parameters proposed to account for the large specific heat jump have
been shown to give the opposite field-dependence for the Bragg peaks
to that observed in experiment \cite{Mason95}.  To date, neither the
small moments (about 1\% of the full uranium
moment\cite{Broholm87,Broholm91}) nor the large specific heat
jump\cite{Palstra} $\Delta C/C\approx 1$ at $T_N$ have been explained.
In this paper, we present a model which may account for the small
moments, and discuss additional requirements needed to attain
sufficiently large specific heat jump $\Delta C$ and linear specific
heat coefficient $\gamma$.

The sharp, dispersive spin excitations in \urs\ are longitudinal, and
are well described in terms of a transition between two crystal-field
singlets\cite{Broholm87,Broholm91}; no sharp higher energy levels have
been observed, consistent with the highly anisotropic susceptibility.
The low transverse susceptibility ($\chi^z/\chi^{x,y} \sim 5$ as $T\to
0$) indicates that the next higher degenerate crystal-field levels are
at energies an order of magnitude higher than the $\Delta=2.4$~THz of
the first excited singlet state.  A broad longitudinal continuum has
been observed at higher energies, but we focus here on the sharp
crystal-field-like features of the spectrum.  Although a doublet
crystal-field ground state has been proposed\cite{Amitsuka} for U
diluted in ThRu$_2$Si$_2$ with its different cell parameters, we
follow the evidence from neutron scattering that \urs\ has two
low-lying singlets separated by a gap $\Delta$.

In an attempt to describe the magnetism of \urs, we consider the
following Hamiltonian which is an Ising-Kondo lattice model with
transverse field:
\begin{equation}
H=\sum_{\svec{k}}\epsk\gdg\psi\dg_{\svec{k}}\psi\gdg_{\svec{k}}
   +I\sum_i S^z_i \psi\dg_i\hlf\sigma^z\psi\gdg_i
   -\Delta\sum_i S^x_i
\label{Hamiltonian}
\end{equation}                 
where $\psi\dg_i\equiv(c\dg_{i\up},c\dg_{i\dn})$ (with
$\psi\dg_{\svec{k}}$ as its Fourier transform), and $c\dg_{i\sigma}$
creates a conduction electron of spin $\sigma=\up,\dn$ at lattice site
$i$; $\epsk$ is the conduction band energy, with width $2W$.  The
$S=1/2$ operators $\vec{S}_i$, normalized so $\vec S_i^2=3/4$,
correspond to the two-level system consisting of the two low-lying
singlets which are separated by an energy $\Delta$, and we thus regard
it as the spin operator for $f$ electrons:  the isotropic Kondo
exchange $I\sum_i\vec{S}_i\cdot(\psi\dg_i\hlf\bvec{\sigma}\psi\gdg_i)$
reduces to the above Ising form when the true $f$-spin is projected
onto the lowest two singlet states.  Details about the crystal-field
splitting are given in Appendix~\ref{app:crystal-field}.

In \refcite{Mason-Buyers91}, parameters of this Hamiltonian were
estimated that agreed with the expected 6$d$ conduction-electron
bandwidth $2W\approx 500$~THz, on-site exchange $I\approx 25$~THz
comparable to that in other heavy-fermion systems, and the
singlet-singlet gap $\Delta\approx 2.4$~THz observed in neutron
scattering at low T.  These authors also reported an
order-of-magnitude mass enhancement from spin fluctuations above
$T_N$, but as discussed in Sec.~\ref{sec:effmass}, a lowest order
perturbative calculation shows that the mass enhancement is small
unless I/W is $O(1)$.

While \urs\ does not necessarily correspond exactly to half-filling,
in this paper we restrict attention to the half-filled case, to
one-dimensional (1D) and centred tetragonal\cite{note1} ({\sc ct})
lattices, and a conduction-electron band constructed from
inter-sublattice hopping:
\begin{equation}
\sum_{\svec{k}}\epsk\gdg\psi\dg_{\svec{k}}\psi\gdg_{\svec{k}}
=-t\sum_{\nnij,\sigma} c_{i\sigma}\dg c_{j\sigma}\gdg+\hc
\end{equation}
where $\nnij$ denotes bonds between nearest neighbours on opposite
sublattices.  Specifically,
\begin{equation}
\epsk=\left\{ \begin{array}{ll} 
      -2t\cos(ka), & \mbox{1D} \\
      -8t\cos(k_xa/2)\cos(k_ya/2)\cos(k_zc/2), & \mbox{{\sc ct}}
              \end{array} \right.
\end{equation}
so that the bandwidth $2W$ is $4t$ ($16t$) in the 1D ({\sc ct})
lattice.  Here the conventional {\sc ct} unit cell, considering the
{\sc ct} lattice to be {\sc bct}, measures $a\times a\times c$.

This paper proceeds as follows.  We study the model mainly in the
mean-field approximation, and find an extensive low-moment regime as
well as reasonable N\'eel temperature, although the specific heat,
while showing a jump at $T_N$, is not of sufficient magnitude.  Using
a single-impurity model for the specific heat, we speculate on ways to
obtain a large specific heat coefficient $\gamma$.  We also find that
in the limits of small bandwidth and low coupling, direct comparisons
with other electron correlation models are possible.

\section{Mean-Field Approximation}

We present mean-field results for our Hamiltonian which
naturally provide a region of parameter space in which
the required small moments occur.

Taking as our mean-field {\it ansatz\/} 
$S^z_j=(-1)^j m_f  + \delta S^z_j$
for the $f$-electron spins and
$\psi\dg_j\hlf\sigma^z\psi\gdg_j\equiv s^z_j
=(-1)^{j+1} m_c+\delta s^z_j$
for the conduction-electron spins, where
$(-1)^j\equiv\exp(i\vec{Q}\cdot\vec{R}_j)$ alternates between opposite
sublattices, our Hamiltonian becomes, upon dropping the product of
fluctuations,
\begin{eqnarray}
H_{\rm MF} &=& \sum_{\svec{k}}
(\psi\dg_{\svec{k}},\psi\dg_{\svec{k}+\svec{Q}})
\!\left(\!
  \begin{array}{cc}
  \epsilon_{\svec{k}} & Im_f\hlf\sigma^z \\
  Im_f\hlf\sigma^z    & \epsilon_{\svec{k}+\svec{Q}}
  \end{array}
\!\right)
\!\left(\!
  \begin{array}{c}
  \psi\gdg_{\svec{k}}\\
  \psi\gdg_{\svec{k}+\svec{Q}}
  \end{array}
\!\right)\nonumber\\
&&-\sum_j [ (-1)^j Im_c S_j^z+\Delta S_j^x ]
 +{\cal N} I m_f m_c \:\:,
\end{eqnarray}
where ${\cal N}$ is the number of uranium atoms, and $\vec{k}$ runs
over the magnetic Brillouin zone.  Minimizing the free energy, we
obtain self-consistent mean-field equations for the $f$- and
conduction-electron moments $m_f$ and $m_c$, given here for bands
$\epsilon_{\svec{k}}$ satisfying the generalized nesting condition
$\epsk=-\epsilon_{\svec{k}+\svec{Q}}$ for all $\vec{k}$ (which occurs,
for example, for nearest-neighbour inter-sublattice hopping in 1D,
square, and {\sc ct} lattices):
\begin{eqnarray}
m_f & = &
  {Im_c \over 2\tilde{\Delta} } \tanh(\beta\tilde\Delta/2) \:\:,\\
m_c&=& {Im_f \over {\cal N}} \sum_{\svec{k}}
 \frac{\tanh(\beta\tepsk/2)}{2\tepsk} \:\:,
\end{eqnarray}
where $\pm\tepsk\gdg\equiv\pm\sqrt{\epsk^2+(\hlf Im_f)^2}$ and
$\tilde{\Delta}\equiv\sqrt{\Delta^2+(Im_c)^2}$ are the mean-field
energy levels, and $\beta$ is the inverse temperature.  We note that
the mean-field moments $m_f$ and $m_c$ both have mean-field critical
exponent ${1\over2}$ and are therefore proportional to each other near
the N\'eel temperature $T_N$, which is obtained by the solution of
\begin{equation}
\coth(\Delta/2k_B T_N)
={I^2\over{\cal N}}
\sum_{\svec{k}}
{\tanh(\epsk/2k_B T_N)\over 4\Delta\epsk}\:\:.
\end{equation}
The specific heat per uranium atom is
\begin{eqnarray}
C &=& \frac{k_B}{{\cal N}}
      \sum_{\svec{k}}
   \left[(\beta\tepsk)^2                 
   -  \frac{I^2}{k_B^2 8T}{\partial m_f^2\over\partial T}  \right] 
      \sech^2(\beta\tepsk/2) \nonumber\\
  &&+ {k_B\over 4}
    \left[(\beta\tilde{\Delta})^2
      - \frac{I^2}{2k_B^2 T}{\partial m_c^2\over\partial T}  \right]
      \sech^2(\beta\tilde{\Delta}/2)\:\:.
\end{eqnarray}          

For comparison to experiment, we note that a staggered magnetization
$m_f=0.5$ in our $S_f=\hlf$ model corresponds to a true moment of
1.2$\mu_B$, as this is the observed matrix element of the true $S^z$
operator between the two singlets\cite{Broholm91}.  In
Fig.~\ref{mf-1D-ct-vary-parameters} we show that small $f$-moments
naturally arise from the mean-field approximation in a region of
parameter space not too far from the values of
\refcite{Mason-Buyers91} quoted above.  Mean-field theory also gives a
N\'eel temperature of appropriate order.  We expect that in a full
calculation fluctuation effects would reduce both the moments and
N\'eel temperature.  Thus the values of the parameters could differ
from those derived by matching mean-field theory to experiment.

It is difficult within mean-field theory to obtain both the N\'eel
temperature and moments of experiment since we do not have freedom to
simply scale all parameters: the singlet-singlet spacing cannot be
very different from $\Delta=2.4$~THz.

Since the Fermi surface is perfectly nested (in the generalized sense
defined above) by the antiferromagnetic wave-vector --- this is true
for nearest-neighbour inter-sublattice hopping in 1D, square, and {\sc
ct} lattices --- we find that, as expected, the conduction electrons
exhibit logarithmic response to the field produced by the $f$-system:
if $Im_f\ll W$, then $m_c\sim m_f |\log m_f|^d$ in $d$ dimensions.
This behaviour is caused by the sharp corners found in the Fermi
surface only at half-filling, and is therefore not present away from
half-filling, or for more general Fermi surfaces.

The temperature-dependence of the $f$-moment and specific heat for
selected parameters are given in Fig.~\ref{mf-and-C-vary-T} for the 1D
case, showing a jump $\Delta C/C\approx 15$\% with $C$ too small by a
factor of 6 to fit experiment.  The jump arises mainly from the
formation of a spin gap $Im_f$ in the conduction electron spectrum.

\section{Analytic Behaviour in Two Limits}

We now study our Hamiltonian in two relatively simple limits, namely
those of small bandwidth and small coupling which to first order lead,
respectively, to the Heisenberg model with Ising anistropy and the
Hubbard model.  It is clear that the inclusion of additional terms in
our Hamiltonian to model the real system will dominate some of the
terms we find in the effective Hamiltonians in these limits.  For
example, had we included a nearest-neighbour Coulomb interaction
between conduction electrons, the $U$ term we obtain for the Hubbard
model in Sec.~\ref{sec:smallI} below would be negligible\cite{Rice}.
So while we do not believe that these limits pertain to the real
material \urs, we study them to obtain a more global picture of our
Hamiltonian and hopefully understand more about its behaviour in the
real, relatively complicated, parameter regime.
Fig.~\ref{fig:phasediag} depicts a summary of the results of this
section.

\subsection{Small Bandwidth Limit}

If $t=0$, then our Hilbert space breaks up into an independent-site
description which is easily diagonalized to give the degenerate
energy-level diagram for each site shown in Fig.~\ref{fig:levels}.  In
the small $t$ limit, we are restricted to the ground state doublet in
Fig.~\ref{fig:levels}, in the sector of the Hilbert space in which
there is exactly one conduction electron per site.  Taking the
standard strong-coupling approach for
\begin{equation} 
t\ll\qrtr\sqrt{I^2+4\Delta^2}-\hlf\Delta,
\end{equation}
we obtain in second-order degenerate perturbation theory (similar to
that for the large-$U$ Hubbard model\cite{AffleckLH}) the Heisenberg
Hamiltonian with Ising anisotropy:
\begin{equation}
H_{\rm eff} = \sum_{\nnij,a} J^a s^a_i s^a_j
\end{equation}
where
\begin{eqnarray}
J^{x,y}&=&\frac{8t^2}{I^2(I^2+4\Delta^2)^{3/2}}\left(
        10I^2\Delta^2 + 32\Delta^4 \right)\\
J^z&    =&\frac{8t^2}{I^2(I^2+4\Delta^2)^{3/2}}\left(
  I^4 + 10I^2\Delta^2 + 32\Delta^4 \right)
\end{eqnarray}
In the limit $\Delta/I\to 0$ we have $J^{x,y}\to 0$, which gives the
pure Ising Hamiltonian.   The opposite limit $I/\Delta\to 0$ results
in the Heisenberg Hamiltonian with
\begin{equation}
J^{x,y,z}\to {32t^2\Delta\over I^2} \label{I.ll.Delta}
\end{equation}
and Ising anisotropy 
\begin{equation}
J^z-J^x\to {t^2I^2\over\Delta^3}\to 0.  \label{Ising.anis}
\end{equation}

\subsection{Low Coupling Limit}
\label{sec:smallI}

Using a path-integral method, we integrate out the localized $f$-spins
in the limit $I\ll\Delta$, and we find the imaginary-time effective
Lagrangian for the conduction electrons to be
\begin{equation}
{\cal L}_{\rm eff}^I = \sum_{\svec{k}\sigma}
    c\dg_{\svec{k}\sigma}\left( \partial_\tau+\epsk\right)c_{\svec{k}\sigma}
    - \frac{I^2}{4\Delta}\sum_j
          \left(s^z_j(\tau)\right)^2
    + \frac{I^2}{4\Delta^3}\sum_j
          \left(\partial_\tau s_j^z(\tau)\right)^2
    + \cdots
\end{equation}
where $\cdots$ represents terms of higher order in derivatives and in
$I/\Delta$.

With just the first correction term (the one proportional to
$I^2/\Delta$), noting that
$(s^z_j)^2=\frac{1}{4}n_j-\frac{1}{2}n_{j\up}n_{j\dn}$, we have an
effective Hubbard Hamiltonian with
\begin{equation}
U=\frac{I^2}{8\Delta}
\end{equation}
which in the strong-coupling limit $t\ll U$ gives a $t$--$J$ model
with
\begin{equation}
J=\frac{4t^2}{U}=\frac{32t^2\Delta}{I^2}
\end{equation}
which agrees with our small $t$ limit in the special case of
$I/\Delta\to 0$ shown in \eq{I.ll.Delta}.

At half-filling, the Hubbard model is believed, on the basis of
mean-field theory, to exhibit a Mott-Hubbard transition at
infinitesimal $U$ to an antiferromagnetic phase\cite{Fradkin}.  In $d$
dimensions, the staggered magnetization is
\begin{equation}
m_c\propto \frac{t}{U}\exp\left[-(at/U)^{1/d}\right],
\end{equation}
with the singularity arising from the Fermi surface
corners\cite{AffleckLH}.  For the model of current interest in the
($d=3$) {\sc ct} lattice, we determine the coefficient in the
exponential to be $a=3\pi^3/2$.

We now interpret 
\begin{equation}
\partial_\tau s_j^z=[H,s_j^z]\approx
 [H_0,s_j^z]+O(I/\Delta),
\end{equation}
where $H_0$ is the Hamiltonian with $I=0$.  This results in an
effective Hamiltonian which has interaction terms
{{{
\begin{eqnarray}
H_{\rm int} &=& -\frac{I^2}{4\Delta}\sum_j (s^z_j)^2
-\frac{t^2 I^2}{2\Delta^3}
\left\{
 2\sum_{\nnij}\left(
s^+_i s^-_j+c\dg_{i\up}c\dg_{i\dn}c_{j\dn}\gdg c_{j\up}\gdg
+\hc
-\sum_{\sigma} n_{i\sigma}n_{j\sigma}\right)\right.\nonumber\\
&&\left.-\sum_{\nn{ijk}}\left(
      c\dg_{j\up}c\dg_{j\dn}(c_{i\dn}\gdg c_{k\up}\gdg
     +c_{k\dn}\gdg c_{i\up}\gdg )
+c\dg_{j\up}c_{j\dn}\gdg (c\dg_{i\dn}c_{k\up}\gdg
                             +c\dg_{k\dn}c_{i\up}\gdg)
\sum_\sigma n_{j\sigma}c\dg_{i\sigma}c_{k\sigma}\gdg 
+\hc\right) \right\}.
\end{eqnarray}
where $\sum_{\nn{ijk}}\equiv\sum_{j(i=j-1, k=j+1)}$ (valid only in 1D)
and the hermitian conjugate $\hc$ applies to all terms to the left
within the brackets.

We note that in the small $t$ limit, in which we project onto the
singly-occupied subspace, all but the first two terms disappear and
the result agrees precisely with the Ising anisotropy found in
\eq{Ising.anis}.

\section{Large Effective Mass} 
\label{sec:effmass}

While the mean-field approximation gives experimentally interesting
(and hopefully accurate) values for the zero-temperature moment and
N\'eel temperature, in addition to not predicting a large specific
heat jump, it does not account for another fundamental property of
\urs: the large value of $\gamma \approx 180$~mJ/mol-K$^2$, the
zero-temperature intercept of $C/T$ above $T_N$, {\it i.e.} the
heavy-fermion mass.  This is because, once the moments have gone to
zero, the mean-field specific heat simply becomes that of free
conduction electrons plus a Schottky term from the $f$-moments.

One possible way of improving on mean-field theory above $T_N$ is to
extrapolate known or conjectured results on the single-moment version
of the problem to the lattice version.  That is, we consider the
Hamiltonian of \eq{Hamiltonian} with a single spin, located at the
origin.  This model may be recognized as a particular realization of
the well-studied problem of a spin coupled to a heat bath.  Assuming a
spherically symmetric dispersion relation, and expanding the fermion
fields in spherical harmonics, only one harmonic interacts with the
impurity spin so the problem is effectively
one-dimensional\cite{note2}.  This one-dimensional spin-fermion
problem may be bosonized at low energies, giving the spin-boson
problem.  This model has been used to study the effect of dissipation
on tunneling.  The dimensionless strength of the dissipation is
measured by $I/W$; $\Delta$ corresponds to the tunneling matrix
element.  Since we are apparently in the weak dissipation limit,
$I/W\ll 1$, the spin has a unique ground state with $\langle
S^x\rangle\neq 0$. (At stronger dissipation there are two ground
states with $\langle S^z\rangle=\pm m\neq 0$.)  The spin-boson problem
in turn is equivalent, at low energies, to the Kondo problem.  The
weak dissipation case corresponds to antiferromagnetic Kondo coupling
with a screened ground state.

According to \refcite{Leggett}, this connection with the spin
dissipation problem gives $\delta\gamma\propto f(I/W)/\Delta$ with
$f(x)\to x^2$ as $x\to 0$ and $f(1)\approx 1$.  Using perturbation
theory to second order in $I/W$, we indeed find that 
\begin{equation}
\gamma=\gamma_0+\frac{\pi^2I^2\rho^2_0}{12\Delta}+\cdots,
\end{equation}
where $\rho_0=1/2W$ is the density of states at the Fermi level.  But
we need $\gamma\approx \Delta^{-1}$ to fit experiment; with
$I/W\approx .1$ this is two orders of magnitude too small.  In order
to obtain a large enough $\gamma$ we would apparently have to choose
$I/W\approx 1$, in which case the moment would not be small according
to \figref{mf-1D-ct-vary-parameters}.

Thus, making the non-interacting impurity approximation, we cannot
explain the large $\gamma$ value without assuming a large value of
$I/W$.  It is possible that treating our model more accurately,
including {\sc rkky} interactions, will lead to a sufficiently large
$\gamma$.  Alternatively, the model may be missing some important
physics.  Recall that we have thrown away the uranium crystal field
levels which are an order of magnitude higher; reinstating them brings
the spin-flip part of the Kondo interaction back into play.  

The huge mass enhancement of the charge carriers in many heavy-fermion
compounds is often explained in terms of the Kondo effect. Besides the
bare conduction-electron bandwidth, the Kondo effect introduces a much
smaller energy scale, namely the Kondo temperature $T_K$, below which
the local moment degrees of freedom are frozen.  The same small energy
scale also gives rise to an effective narrow band with an enhanced
density of states:  if the Kondo temperature is of order $\Delta$ (the
crystal-field splitting) or larger, the spin-flip parts of the Kondo
coupling can renormalize to the strong-coupling fixed point producing
a $\gamma$ per ion of $O(1/T_K)$.  In \urs, we expect this
renormalization to be cut off at a scale of order $\Delta$; however,
one still expects the same kind of Kondo screening process over a wide
range of energy scales from the order of the bare bandwidth down to
$O(\Delta)$.  While this effect is clearly not included when we write
down the low-energy effective Ising-Kondo lattice model, a
semi-phenomenological way of treating it is to use our model with a
greatly reduced effective bandwidth of $O(T_K)$.  This effective
low-energy theory is valid at low energy scales after the higher
crystal-field levels have been integrated out and the associated Kondo
effect has been taken into account.

\section{Discussion}

We have shown that within the mean-field approximation, small moments
are predicted for a range of parameters because of a logarithmic
response of the conduction electrons at half-filling.  A specific heat
jump is obtained mainly from the formation of a conduction-electron
spin gap at $T_N$.

The question remains as to the nature of the dynamical narrowing of
the bandwidth.

We have given comparisons of our model in various limits to other
models, such as the Hubbard, Heisenberg, and Ising models.  Because
anisotropy is maintained in these limits, we expect that the
one-dimensional system will still order at zero temperature.

\acknowledgments{
This research is supported in part by {\sc nserc} of Canada;
{\sc wjlb} is also grateful to {\sc ubc},
{\sc ciar}, and {\sc aecl} for support during a sabbatical visit. 
{\sc aes} gratefully acknowledges support from the Izaak Walton Killam
Memorial Foundation.}

\begin{appendix}
\section{Details of the Crystal-Field Splitting in \urs}
\label{app:crystal-field}

We take the uranium ion to have total angular momentum $J=4$, and the
crystal-field Hamiltonian which splits the 9-fold degeneracy has the
form\cite{Amoretti}
\begin{equation}
H_{\rm CF}=
\HCFterm{2}{0}+
\HCFterm{4}{0}+
\HCFterm{4}{4}+
\HCFterm{6}{0}+
\HCFterm{6}{4}
\end{equation}
where $\StevensOp{m}{n}$ are the Steven's angular momentum operators
for centred tetragonal symmetry; $\StevensCoeff{m}{n}$ are the
coefficients for \urs.  With the coefficients we choose in an attempt
to match the neutron-scattering experiments, this Hamiltonian results
in two low-lying singlets as well as two high-lying doublets and three
singlets.  We represent what we consider the important structure in
Fig.~\ref{fig:CFlevels}.  The two low-lying singlets are
\begin{eqnarray}
\ket{0}&=&\Gam{1}{t1}\equiv\epsilon(\ket{4,4}+\ket{4,-4})+\gamma\ket{4,0},\\
\ket{1}&=&\Gam{}{t2}\equiv\frac{1}{\sqrt{2}}(\ket{4,4}-\ket{4,-4})
\end{eqnarray}
where $\ket{J,m_J}$ are the states of total angular momentum $J$ and
azimuthal quantum number $m_J$.  The operators $J^{\pm}$ bring
$\ket{0}$ and $\ket{1}$ into the high-energy states, while
$\bra{0}J^z\ket{1}=8\epsilon/\sqrt{2}$.  Writing $\vec{s}$ for the
conduction electron spin, the Kondo interaction
$\vec{J}\cdot\vec{s}=J^zs^z+\hlf(J^+s^-+J^-s^+)$ reduces to just its
Ising part $J^zs^z$ when considering only the two low-lying singlets.
Now we define a dimensionless and normalized spin-$\hlf$ operator
$S^z$ such that $J^z=(8\epsilon/\sqrt{2})S^z$.  In order to implement
the separation $\Delta$ between the two low-lying singlets, we
introduce an operator $S^x$ which is diagonal in this subspace:
\begin{eqnarray}
S^x\ket{0}&=&-\hlf\ket{0};\\
S^x\ket{1}&=&+\hlf\ket{1}.
\end{eqnarray}
Thus we obtain the terms $I\sum_iS^z_is^z_i+\Delta\sum_i S^x_i$ in our
Hamiltonian (\ref{Hamiltonian}).
\end{appendix}

\begin{figure}
\centerline{\epsfxsize=19.25 cm \epsffile{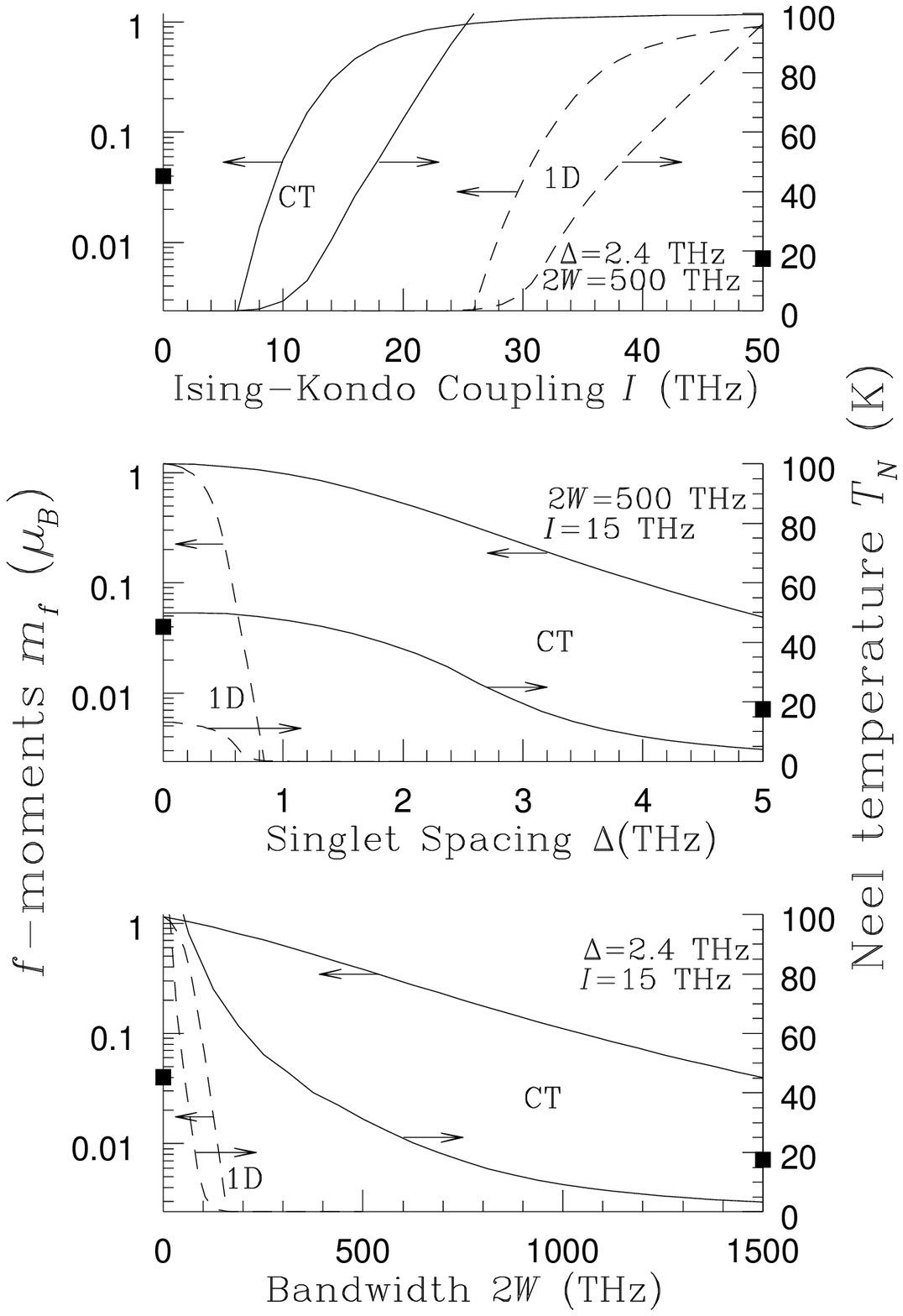}}
\caption{Zero-temperature $f$-moments and N\'eel temperatures for 1D
(dashed lines) and {\sc ct} (solid lines) lattices as functions of the
bandwidth, singlet spacing, and Ising-Kondo coupling.  In each case we
fix two of the parameters and vary the other.  Experimental values
$m_f=0.04\mu_B$ and $T_N=17.5$~K are indicated on the axes.}
\label{mf-1D-ct-vary-parameters}
\end{figure}
 
\pagebreak

\begin{figure}
\centerline{\epsfxsize=20 cm \epsffile{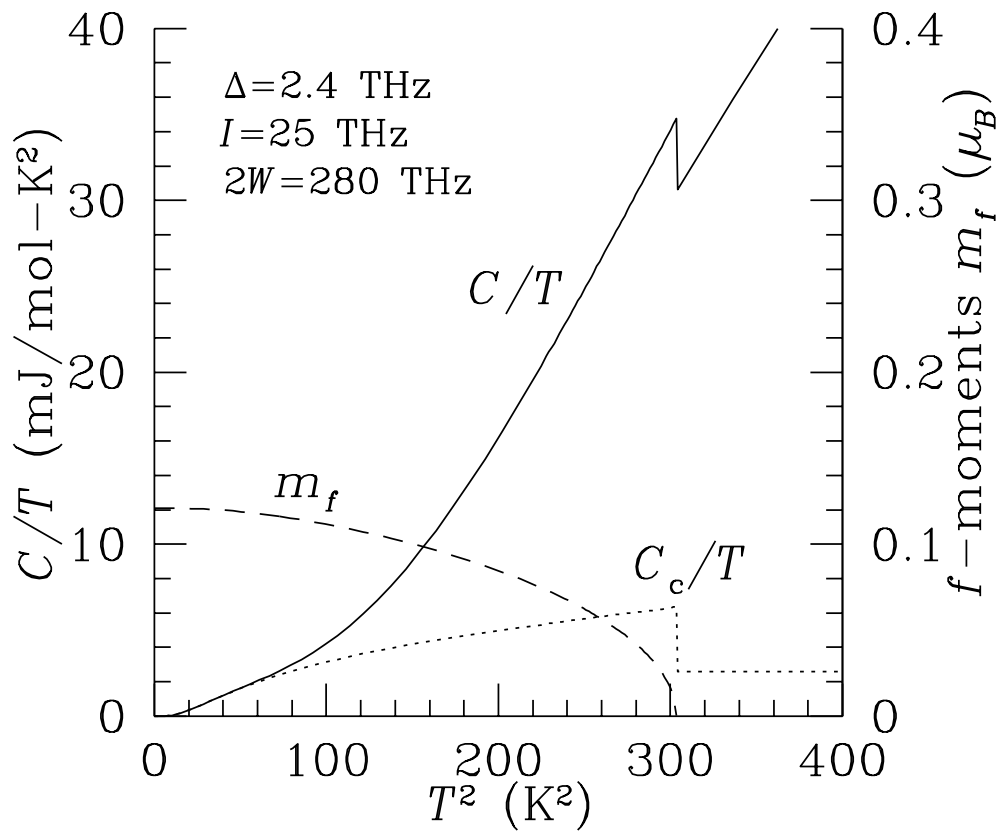}}
\caption{Selecting parameters
for the 1D lattice, we plot $f$-moments 
and the specific heat divided by temperature against $T^2$.
The conduction-electron part, $C_c/T$, is the dotted line.}
\label{mf-and-C-vary-T}
\end{figure}

\pagebreak

\begin{figure}
\centerline{\epsfxsize=12 cm \epsffile{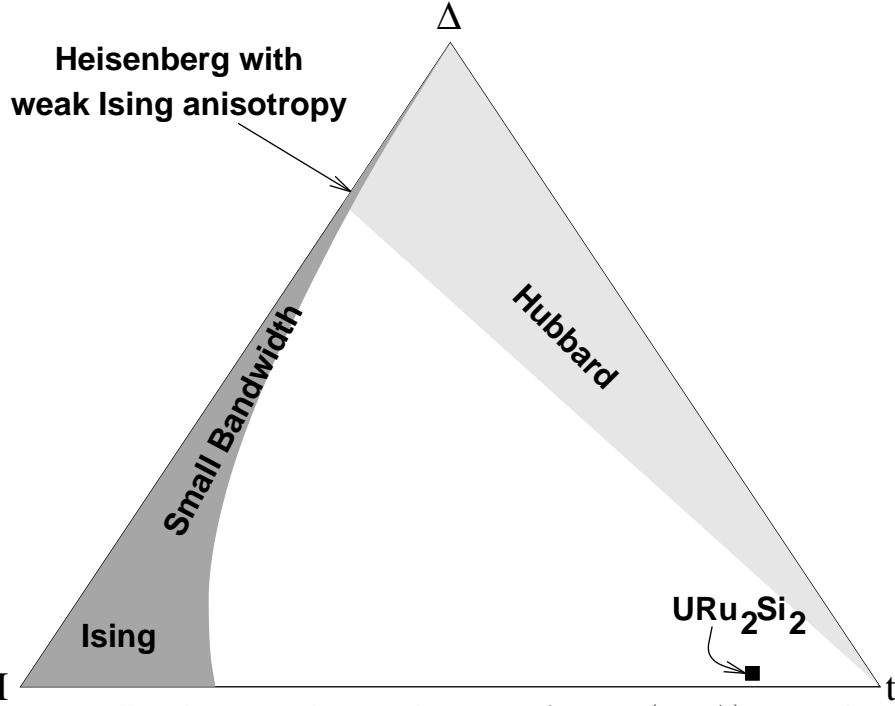}}
\caption{Since overall scaling is irrelevant, the space of points
$(t,I,\Delta)$ is two-dimensional.  Here the points are represented in
such a way that $t:I:\Delta=d_{I\Delta}:d_{t\Delta}:d_{tI}$ where
$d_{ab}$ is the distance from the point to the line joining $a$ and
$b$.  The regions we have examined are the shaded ones.  (For the
purposes of this figure, we interpret $a\ll b$ as $a<b/3$.)}
\label{fig:phasediag}
\end{figure}

\pagebreak

\begin{figure}
\begin{center}\begin{tabular}{cc@{\hspace{2cm}}ll}
& Energy & States\\
\Elevel & $+\qrtr\sqrt{I^2+4\Delta^2}$ 
 & $ \ket{\!\up}\ket{\!\nearrow}$~, &
   $ \ket{\!\dn}\ket{\!\searrow}$  \vspace{3mm}\\
\Elevel & $+\hlf\Delta$ & $\ket{0}\ket{\!\rightarrow}$~, &
                          $\updn\ket{\!\rightarrow} $ \\
\Elevel & $-\hlf\Delta$ & $\ket{0}\ket{\!\leftarrow}$~, &
                          $\updn\ket{\!\leftarrow}$  \vspace{3mm}\\
\Elevel & $-\qrtr\sqrt{I^2+4\Delta^2}$ 
  & $ \ket{\!\up}\ket{\!\swarrow}$~, &
    $ \ket{\!\dn}\ket{\!\nwarrow}$
\end{tabular}\end{center}
\caption{Spectrum of the independent-site Hamiltonian for $t=0$.  Here
the first ket represents the conduction electrons:
$\ket{\sigma}=c_\sigma\dg\ket{0}$ for $\sigma=\up,\dn$; $\updn=
c_\up\dg c_\dn\dg\ket{0}$; $\ket{0}$ is the ``vacuum''.  The second
ket represents the state of the localized $f$ electron:
$\ket{\!\nearrow}=\ket{\eta_{\theta}};
\ket{\!\searrow}=\ket{\eta_{\pi-\theta}};
\ket{\!\rightarrow}=\ket{\eta_{\pi/2}};
\ket{\!\leftarrow}=\ket{\eta_{-\pi/2}};
\ket{\!\nwarrow}=\ket{\eta_{-\theta}};
\ket{\!\swarrow}=\ket{\eta_{\theta-\pi}}$, and
$\ket{\eta_{2\phi}}=(f_\up\dg\cos\phi+f_\dn\dg\sin\phi)\ket{0}$ is the
state in which the $f$-electron spin lying in the $x$-$z$ plane makes
an angle $2\phi$ with the positive $z$-axis measured toward the
positive $x$-axis --- note that
$\bra{\eta_{2\phi}}\eta_{2\phi'}\rangle=\cos(\phi-\phi')$.  It is easy
to see that here $\theta=\tan^{-1} (2\Delta/I)$ is the tilt produced
by the ``transverse magnetic field'' $\Delta$.}
\label{fig:levels}
\end{figure}

\pagebreak

\begin{figure}
\centerline{\epsfxsize=8 cm \epsffile{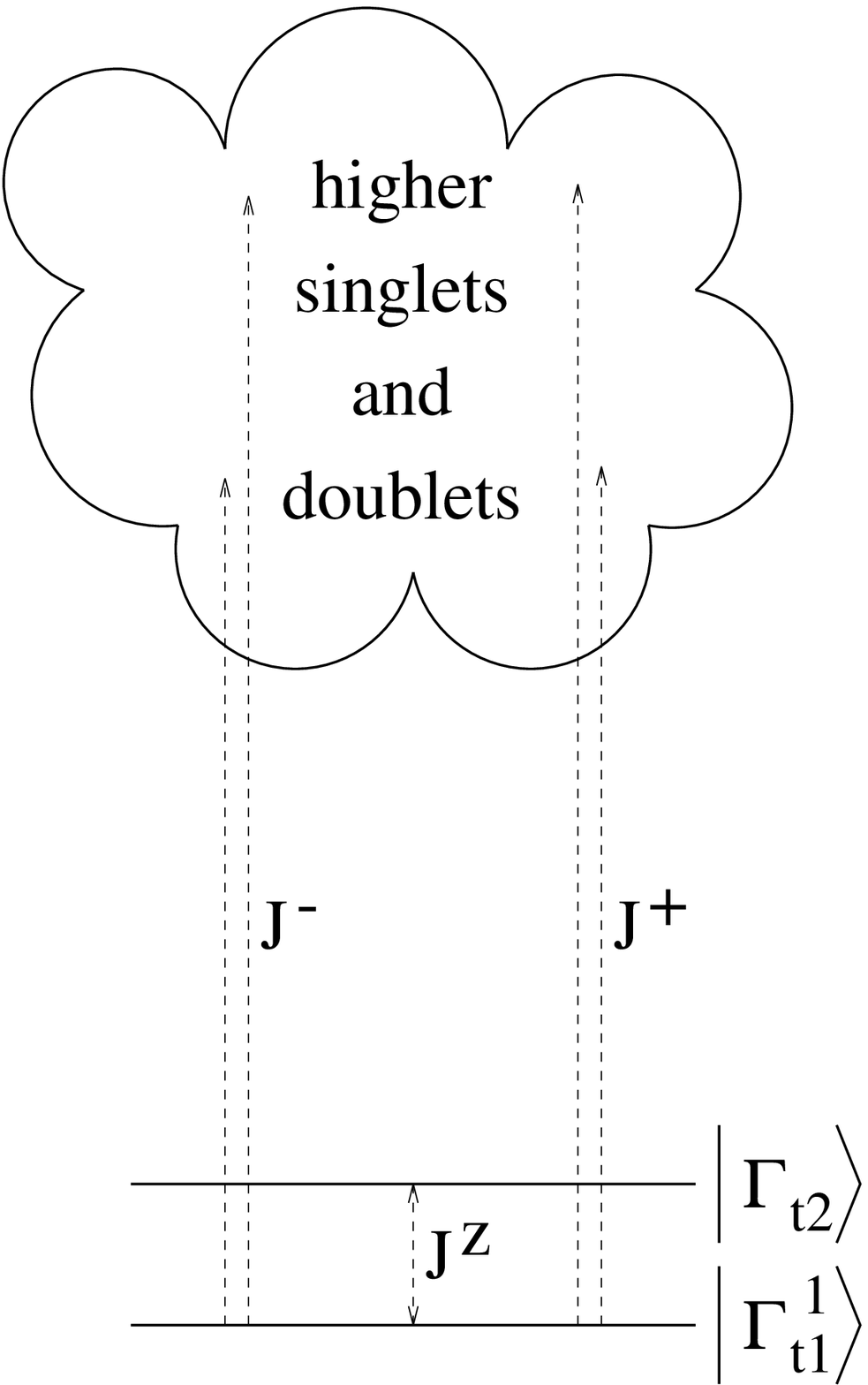}}
\caption{Spectrum of the crystal-field Hamiltonian, showing states
related to the low-lying singlets by application of the operator
$\vec{J}$.}
\label{fig:CFlevels}
\end{figure}


\begin{references}

\bibitem{Broholm87}
C.~Broholm,
J.K.~Kjems, W.J.L.~Buyers, P.~Matthews,
T.T.M.~Palstra, A.A.~Menovsky, \& J.A.~Mydosh,
{\sl Phys. Rev. Lett.} {\bf 58} (1987) 1467.

\bibitem{Broholm91}
C.~Broholm, 
H.~Lin,
P.T.~Matthews, T.E.~Mason, W.J.L.~Buyers, M.F.~Collins,
A.A.~Menovsky, J.A.~Mydosh, \& J.K.~Kjems,
{\sl Phys. Rev. B} {\bf 43} (1991) 12809.

\bibitem{Ramirez}
A.P.~Ramirez, P.~Coleman, P.~Chandra, E.~Br\"uck, A.A.~Menovsky,
Z.~Fisk, \& E.~Bucher,
{\sl Phys. Rev. Lett.} {\bf 68} (1992) 2680.

\bibitem{Barzykin}
V.~Barzykin \& L.P.~Gor'kov,
{\sl Phys. Rev. Lett.} {\bf 70} (1993) 2479.

\bibitem{Cox}
D.L.~Cox, {\sl Phys. Rev. Lett.} {\bf 59} (1987) 1240.

\bibitem{Walker93}
M.B.~Walker, W.J.L.~Buyers, Z.~Tun, W.~Que, A.A.~Menovsky, 
  \& J.D.~Garrett,
  {\sl Phys. Rev. Lett.} {\bf 71} (1993) 2630;
W.J.L.~Buyers, Z.~Tun, T.~Petersen, T.E.~Mason, J.-G.~Lussier,
  B.D.~Gaulin, \& A.A.~Menovsky,
  {\sl Physica B} {\bf 199\&200} (1994) 95;
M.B.~Walker, Z.~Tun, W.J.L.~Buyers,  A.A.~Menovsky, \& W.~Que,
  {\sl Physica B} {\bf 199\&200} (1994) 165.

\bibitem{Mason95}
T.E.~Mason, W.J.L.~Buyers, T.~Petersen, A.A.~Menovsky, 
  \& J.D.~Garrett,
  {\sl J. Phys.: Cond. Mat.} {\bf 7} (1995) 5089.

\bibitem{Palstra}
T.T.M.~Palstra, A.A.~Menovsky, J.~van~den~Berg,
A.J.~Dirkmaat, P.H.~Kes, G.J.~Nieuwenhuys, \& J.A.~Mydosh,
{\sl Phys. Rev. Lett.} {\bf 55} (1985) 2727.

\bibitem{Amitsuka}
H.~Amitsuka,
T.~Hidano, T.~Honma, H.~Mitamura, \& T.~Sakakibara,
{\sl Physica B} {\bf 186-188} (1993) 337.

\bibitem{Mason-Buyers91}
T.E.~Mason \& W.J.L.~Buyers,
{\sl Phys. Rev. B} {\bf 43} (1991) 11471.

\bibitem{note1}
Note that there is only one centred tetragonal lattice; {\sc bct} and
{\sc fct} are identical.

\bibitem{Rice}
We thank T.~Maurice~Rice for bringing this to our attention in a
private discussion.

\bibitem{AffleckLH}
For a review, see I.~Affleck,
in \underline{Fields, Strings and Critical Phenomena},
E.~Br\'ezin and J.~Zinn-Justin, eds.,
Les Houches XLIX (North-Holland, Amsterdam, 1990) p.~563.

\bibitem{Fradkin} See, for example, E.~Fradkin, 
\underline{Field theories of condensed matter systems},
(Redwood City:  Addison-Wesley, 1991), p. 32.

\bibitem{note2}
Alternatively, equal-energy surfaces may be used, removing the
assumption of sphericity; see
I.~Affleck, A.W.W.~Ludwig, \& B.A.~Jones,
{\sl Phys. Rev. B} {\bf 52} (1995) 9528 ({\tt cond-mat/9409100}).

\bibitem{Leggett}
A.J.~Leggett,
S.~Chakravarty, A.T.~Dorsey, M.P.A.~Fisher, A.~Garg, \& W. Zwerger,
{\sl Rev. Mod. Phys.} {\bf 59} (1987) 1.

\bibitem{Amoretti}
G.~Amoretti, A.~Blaise, \& J.~Mulak, 
  {\sl J. Mag. Mag. Mat.} {\bf 42} (1984), 65;
G.~Amoretti, A.~Blaise, R.O.A. Hall, M.J. Mortimer, \& R. Tro\'c,
  {\sl J. Mag. Mag. Mat.} {\bf 53} (1986) 299.

\end{references}
\end{document}